\documentclass{llncs}

\usepackage[hidelinks]{hyperref}
\usepackage{subcaption}
\usepackage{longtable}
\usepackage[binary-units]{siunitx}
\usepackage[T1]{fontenc}
\usepackage{cite}
\usepackage{amsmath}

\usepackage{tikz}
\usetikzlibrary{shapes}
\usetikzlibrary{arrows}

\newcommand\rightfill{\hspace{\fill}\mbox{}\linebreak[0]\hspace*{\fill}}

\begin{document}

\pagestyle{headings}

\title{Interactive Simplifier Tracing and Debugging\\in Isabelle\thanks{The final publication is available at \url{http://link.springer.com.}}}
\titlerunning{Interactive Simplifier Tracing and Debugging in Isabelle}

\author{Lars Hupel}
\authorrunning{Lars Hupel}

\institute{Technische Universit\"at M\"unchen \\
\email{lars.hupel@tum.de}}

\mainmatter
\maketitle

\begin{abstract}
  The Isabelle proof assistant comes equipped with a very powerful tactic for term simplification.
  While tremendously useful, the results of simplifying a term do not always match the user's expectation:
  sometimes, the resulting term is not in the form the user expected, or the simplifier fails to apply a rule.
  We describe a new, interactive tracing facility which offers insight into the hierarchical structure of the simplification with user-defined filtering, memoization and search.
  The new simplifier trace is integrated into the Isabelle/jEdit Prover IDE.

\end{abstract}

\section{Introduction}

\emph{Isabelle} is a generic proof assistant \cite{paulson2002isabelle}.
It comes with some very powerful tactics which are able to discharge large classes of proof goals automatically.
This work is concerned with the rewriting tactic, often called the \emph{simplifier}.
It can be used to rewrite subterms according to a user-definable set of equations, which generally means simplifying a term to a normal form.
These equations can have conditions which are recursively solved by the simplifier itself.
Hence, there can be quite a huge number of steps between the original term and its normal form.
Because of that complex work in the background, it is not obvious to the user how the input gives rise to certain terms in the output.

{
  \looseness = -1
By default, this process is completely opaque:
the only observable effect is---given that the simplification succeeded---the (hopefully) simpler term it produced.
If it failed, only an error message without any indication of the reasons is printed, or it might not even terminate at all.
Currently, there is a tracing facility which can be enabled by the user.
It collects data about the steps the simplifier executed, and prints each of them without any high-level structure.
The resulting trace can easily contain many screenfuls of items which the user has to laboriously search for interesting pieces of information.
This could also easily lead to sluggish GUI behaviour in the Isabelle/jEdit Prover IDE because of the huge amount of content which has to be rendered on the screen.
There are various techniques to alleviate that problem, notably limiting the recursion depth.

}

However, this does not solve the fundamental problem of showing only the interesting parts of the trace to the user.
This paper describes the design of a new tracing and debugging mechanism which can be configured to filter trace messages and render a semantically meaningful representation of the simplification trace (Section \ref{sec:design}).
For example, rewriting the term $a < b \implies 0 < b \implies 0 < c \implies 0 < (c + c) / (b - a)$ already produces over 300 trace messages with the old tracing.
Navigating the linear presentation of the trace is non-trivial.
The new tracing shows a hierarchical view of the trace where the user can hide the traces of uninteresting subexpressions.
The user can simply filter the trace for various keywords, e.g.\ ``ac'' (which usually denotes an associativity/commutativity rule if its name contains this string).
For our example expression, this uncovers that for solving the goal---the term is rewritten to \texttt{True}---the simplifier used commutativity of multiplication (it rewrote $2 \cdot c$ to $c \cdot 2$).

The new tracing also offers interactive features like breakpoints and stepping into or over recursive calls (Section \ref{sec:interaction}), which are well-known from debuggers for imperative programming languages such as \texttt{gdb} \cite{gdb}.
In combination with the filtering, this is very effective in narrowing down causes for failure.

Furthermore, this paper contributes a mechanism for incremental debugging, i.e.\ upon changes in the source code the system continues execution where it left off instead of starting from scratch again (Section \ref{sec:memoization}), which is especially useful during exploratory proof development where assumptions and rewrite rules change frequently.
Suppose that in our example, the precondition $a < b$ were absent.
Upon stepping through the simplification, the user realizes that the precondition is missing, and adds it to the goal.
The simplifier runs again, but the user does not have to step through identical subtraces again; instead, only new parts are shown. 

The tracing has been integrated into the Isabelle/jEdit Prover IDE, which is implemented in Scala \cite{wenzel2012isabelle}.
The interplay between the ML process and the JVM process hosting the prover and the IDE, respectively, works via an underlying protocol, of which only a reasonably high-level interface is exposed \cite[\S 3]{wenzel2012asynchronous}.
The new simplifier trace works on both sides:
Trace messages are produced and pre-processed in Isabelle/ML and formatted and renderered in Isabelle/Scala.
In order to reduce the amount of data sent between the processes, there is a multi-staged and user-configurable message filter.
The implementation also supports parallel simplifier runs.
Although the new tracing produces more metadata than previously and incurs an overhead for the communication between the prover and the IDE, the performance is acceptable for user interaction (Section~\ref{sec:eval}).

We compare our work with existing approaches in SWI-Prolog \cite{fruehwirth2012swi,sterling1986art,clocksin2003programming}, Maude \cite{clavel1996principles,clavel2011maude} and other proof assistants (Section \ref{sec:related}).
Both SWI and Maude offer tracing mechanisms that solve similar problems, e.g.\ conditional rules and failing rewrite (or deduction) steps.
Finally, we evaluate the performance of the system and give prospects for further research (Section \ref{sec:eval}).

\paragraph{Terminology}
A \emph{(trace) message} is a piece of structured information about the current state of the simplifier.
Messages requiring a response (possibly by the user) are called \emph{interactive}.
A \emph{rewrite rule} is a theorem stating an equality.
If a rewrite rule is conditional, its preconditions are (in most situations) solved recursively by the simplifier itself.

\section{Design Principles}
\label{sec:design}

The fundamental idea for tracing and debugging is to instrument the simplifier to collect data and influence the flow.
The previous tracing only does the former:
at certain points in the program flow, it prints messages indicating the current state.
For example, when the simplifier is invoked, the function \verb~trace_term~ is called, which takes a \verb~term~ and a \verb~string~.
The system then prints the fixed string ``Simplifier invoked...'' and the term to the regular output channel, where it appears alongside other status information, e.g.\ the goal state.

In the new implementation, the simplifier has been extended with the possibility to install hook functions.
When the simplifier is invoked, the hook function \verb~invoke~ is called which takes a \verb~term~.
The precise formatting of the message, or if it gets printed at all, is left as an implementation detail to the new module.
The internal details are not observable for the simplifier.

This has several advantages:
First, it reduces code clutter, because trace formatting and output is not mixed with the program logic.
The semantics of the program flow is not lost, since it is clear that a call to the \verb~invoke~ hook denotes a (possibly recursive) invocation of the simplifier.
Lastly, influencing the simplifier becomes easy.
For example, the \verb~apply~ hook which denotes an application of a conditional rewrite rule takes a \emph{continuation}\footnote{Continuation passing is a progamming technique in which a function does not return a value, but rather takes a function as argument and invokes this function with the value it computed. In essence, a function \texttt{'a -> 'b} is transformed into a function \texttt{'a -> ('b -> unit) -> unit}.} as an argument, which allows the tracing to replay the program flow if necessary.
This means that instead of just observing the current state of the simplifier, the user is actually able to interactively manipulate it.
Because of pervasive immutability in Isabelle's internal data structures, the simplifier does not have to track any tracing state explicitly for that case.

\begin{figure}[t]
  \centering
  \includegraphics[width=\linewidth]{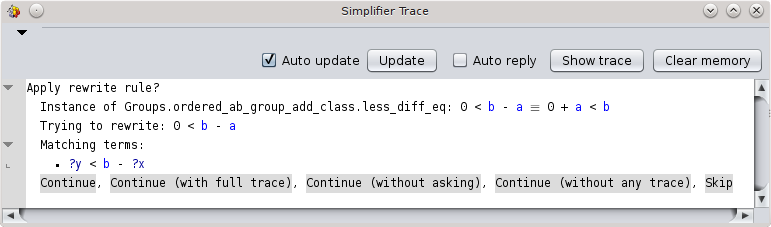}
  \caption{New trace panel in Isabelle/jEdit}
  \label{fig:panel}
\end{figure}

\begin{figure}[t]
  \centering
  \includegraphics[width=\linewidth]{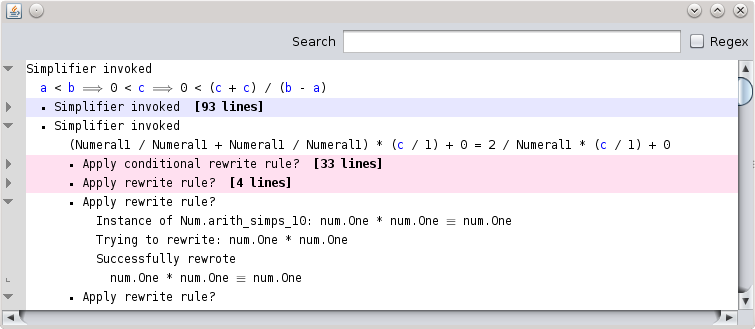}
  \caption{New trace window in Isabelle/jEdit}
  \label{fig:window}
\end{figure}

In the user interface, this is realized by presenting questions to the user at certain points in the program flow (Fig. \ref{fig:panel}).
A question contains the instantiated rewrite rule, the redex, and matching term breakpoints (see Section \ref{sec:settings}).
The simplifier is blocked until the user answers the question, and upon answering, the program flow might deviate from the usual one, e.g.\ certain steps may be skipped.
At any point in time, the user can open a trace window which shows the accumulated trace output (Fig. \ref{fig:window}).
There, the trace can be filtered and sub-parts can be folded for easier navigation.

The new tracing facility is highly configurable and goes to great lengths to keep the number of unwanted messages low.
A related design goal is efficiency, so that the user does not experience long delays.
Together, configurability and efficiency should guarantee that the user interface stays responsive.

\subsection{Hooks and Message Types}
On a high level, these are the types of trace messages the system can send:

\begin{description}
  \item[\texttt{invoke}]
    tells the tracing system that the main entry point of the simplifier has been called.
    This happens on initial invocation of the tactic and on recursive calls when the conditions of rewrite rules are being solved.
  \item[\texttt{apply}]
    guards the application of a (potentially conditional) rewrite rule by the simplifier.
    It is invoked \emph{before} the rule is applied.
    It depends on the user input how the simplification proceeds after that.
    \rightfill\emph{(interactive)}
  \item[\texttt{hint}]
    indicates whether a rewrite step failed or succeeded.
    If it failed, the user is given a chance to inspect the failure, and can decide if the failing step should be tried again (with different settings).
    \rightfill\emph{(possibly interactive)}
  \item[\texttt{ignore}]
    marks a specific part of the trace as obsolete.
    This message is produced when the user wants to retry a failing step.
    It is generated by the tracing system and thus cannot be sent explicitly by the simplifier.
  \item[\texttt{log}]
    emits a log message (with arbitrary content) which will not be further interpreted by the system.
\end{description}

\noindent

Messages may have children.
For example, \verb~apply~ messages are naturally associated with the \verb~invoke~ message emitted in the simplifier invocation.
Hence, each message carries a pointer to the parent message.
The IDE is in turn able to reconstruct the hierarchical structure prior to presenting the full trace output to the user.

Recall the example from the introduction.
When the user calls the simplifier, an \verb~invoke~ message is issued for the whole term.
During simplification, the subterm $0 < c \cdot 2 / (b - a)$ arises.
To simplify it, a rewrite rule with the precondition $0 < x$ is used, where $x$ gets instantiated with $b - a$.
Here, the system first issues a \verb~apply~ message with the name of that rule and its instantiation, and afterwards \verb~invoke~ with the term $0 < b - a$.
This can be solved by the simplifier (after some more rule applications), which yields a \verb~hint(success)~ message, indicating that the original subterm was rewritten to $0 \cdot (b - a) < c \cdot 2$.
The trace window then groups these messages and presents them in one subtree.

\subsection{Settings}
\label{sec:settings}

It is important to determine which questions are relevant.
Showing a question for each step is not feasible, because it might take thousands of steps until a term is rewritten into normal form.
Hence, by default, the system shows no questions at all.
The user is able to specify that behaviour via multiple configuration axes:
Most importantly, \emph{verbosity modes} (which can be changed during a simplification run) and \emph{breakpoints}.

\paragraph{Modes}
The three regular modes of operation are:
\begin{description}
  \item[\texttt{disabled}] do not produce any trace messages at all
  \item[\texttt{normal}] produce messages, but display them only if their contents are triggering a breakpoint
  \item[\texttt{full}] produce messages and display all of them
\end{description}

\noindent
Each of these cases can be combined with a flag which denotes whether the output should just be logged, or presented as a question to the users.
The user can only interactively influence the simplification process in the latter case.

In most cases, it is reasonable to avoid the \texttt{full} mode:
even for seemingly small terms, the potential amount of applied rewrite rules can get quite high.
While the system has no problem producing and transmitting these messages, displaying them takes a while.

\paragraph{Breakpoints}
The user can specify the set of interesting rewrite steps by defining breakpoints.
If a step triggers such a breakpoint, the simplifier is intercepted and the system displays a question. 

In debuggers for imperative languages, the concept of breakpoints is well-known.
Usually, breakpoints are set to lines in the source code, and when the sequential execution of the program hits a marked line, the execution is halted.
Furthermore, many debuggers support conditional breakpoints, where users can specify a condition, and the breakpoint only triggers if that condition is met.

Here, the implementation obviously has to differ from traditional debuggers, because it does not follow a strict sequential execution model.
The principle is easy, though:
each rewrite step in the simplifier consists of a term (the redex) and a theorem (the rewrite rule).
Breakpoints can be set for either of them.
Term breakpoints usually contain patterns and can refer to locally fixed variables.
For example, the breakpoint $\_ > 0$ matches when the term $y + z > 0$ is to be rewritten, where $y$ and $z$ can be any fixed or free variables.
A theorem breakpoint is triggered when the corresponding rewrite rule is used.

Users can declare breakpoints with the usual Isabelle \emph{attribute} mechanism, i.e.\ by adding the string \verb~declare theorem_name[break_thm]~ into the theory sources before the point where the simplifier is invoked (and similarly for term breakpoints).
For example, the panel in Fig. \ref{fig:panel} shows a step which matches a breakpoint declared with \verb~break_term "?y < b - ?x"~ (where \verb~?~ indicates a pattern variable).

\section{User Interaction}
\label{sec:interaction}

User interaction is fundamental to the new tracing.
The system might present the user some questions about the current progress which it deems to be relevant.
The user can then decide how to progress.
In this section, the handling of interactive messages is described.

\subsection{Interactive Messages}

As seen earlier, there are two types of interactive messages which allow the user to influence the outcome of the simplifier:
\texttt{apply} before a simplification step is attempted, and \texttt{hint} for when a simplification step failed.

\paragraph{Message \texttt{apply}}
When a step is attempted, the message contains the instantiated rewrite rule, the redex, and a number of different possible replies.
The user can choose to continue the computation, which instructs the simplifier to apply the specified rule (which requires solving the rule's preconditions first) and thus does not influence the result of the simplifier.
The other option is to skip the current step, even if it would have succeeded.

As a result, the outcome of a simplification run is potentially different from when tracing would be disabled.
Hence, skipping should be used sparingly:
the most common use case would be to find overlapping rewrite rules, i.e.\ multiple rules which match on the same term.

\paragraph{Message \texttt{hint(failed)}}
Often, the user wants feedback immediately when the simplification failed.
Prior to this work, in case of failure the simplifier just does not produce any result, or produces an unwanted result.

On the other hand, with this message type, the new tracing provides insight into the simplification process:
It indicates that the simplifier tried to solve the preconditions of a rewrite rule, but failed.
There are a number of different reasons for that, including that the preconditions do not hold or that a wrong theorem has been produced in the recursive call, which would indicate a programming error in the tactic.
Regardless of the reason, it is possible to \emph{redo} a failed step if (and only if) the original step triggered a user interaction previously.

Consider an example:
The term $f\;t_1$ is to be rewritten.
The rewrite rule $P_1\;x \Longrightarrow f\;x \equiv g\;x$ is applicable and gets instantiated to $P_1\;t_1 \Longrightarrow f\;t_1 \equiv g\;t_1$.
Assume that there is a breakpoint on that particular rule, hence the user is presented a question whether the rule should be applied.
The user chooses to continue, and the simplifier recursively tries to solve the precondition $P_1\;t_1$.
Now assume that this entails application of another conditional rule which does not trigger a breakpoint (hence no question), but this step fails.
In turn, the rewriting of $f\;t_1$ fails.
The system now displays the ``step failed'' message to the user for the outermost failing step.
No messages are displayed for the inner failing steps which caused the outermost one to fail.
This is by design for two reasons:
\begin{itemize}
  \item
    Often, the simplifier has to try multiple rules to prove a precondition.
    This is the case when there are multiple, overlapping rules for a predicate.
    Were the panel to notify the user for each of those steps, this would quickly become very confusing because of a flood of unrelated, and in the end, unimportant failures.
  \item
    If the innermost failure is several layers of recursions away from the original, interesting step, it becomes difficult for the user to establish a causal relationship between the previously answered \texttt{apply} and the subsequent \texttt{hint(failed)} message.
\end{itemize}

Should the user choose to redo the computation, the simplifier state will be rolled back to \emph{before} the last question.
In the above example, the system would ask for the application of the rewrite rule $P_1\;t_1 \Longrightarrow f\;t_1 \equiv g\;t_1$ again.
Of course, answering that question the same way a second time would not change anything.
But it is possible to change the settings and obtain more detailed information.
This requires the simplifier to run anew, which is a consequence of the message filtering (see Section \ref{sec:filtering}).

\subsection{Memoization}
\label{sec:memoization}

Suppose the user invoked the simplifier with a set of rewrite rules and realizes during a simplification run that a rule is missing.
The user then adds the rule to the simplifier and expects not to be asked the same questions they already answered again.
This is similar to debugging an imperative program:
While stepping through the execution, an error is found and the source code is changed accordingly.
After restarting the process, the user reasonably expects to continue at the same point (or a bit earlier) where the debugging session has been exited previously.%
\footnote{In fact, the Java debugger of the \emph{NetBeans} IDE offers a similar feature.
Code changes while debugging can be applied to the running program, which requires reloading the affected classes in the JVM instance.
This completely avoids the problem of reconstructing the original state after restarting the process, because the process is not even being terminated.}
This is not always possible, but as long as editing the source code preserves certain invariants (e.g.\ the arrangement of global variables), resuming at the old execution point is safe.

However, this contradicts the stateless nature of Isabelle's document model:
any changes in the proof text causes the system to discard and reinitialize all active command states.

In the new tracing, a \emph{memoization} system helps in mitigating that issue by trying to reconstruct the original tracing state.
Each time the user answers an \verb~apply~ question, that answer is recorded in a (global) storage.
When the same question appears again later, be it in the same simplification run or in another one, it is automatically answered identically.
All of this happens in the JVM process to avoid cluttering the tracing logic with mutable state.

Redoing a simplification step creates an interesting feature overlap with memoization.
Since fundamentally, redoing a step makes the system display the original question again, a naively implemented cache would auto-answer that question.
As a consequence, the (unchanged) computation would fail again, which would obviously diminish the utility of this question type.
Hence, care has been taken in the implementation to partially clear the cache.

Recall that every time a user chooses to redo a failed computation, the system generates an \verb~ignore~ message.
The memoization mechanism in the IDE picks up the parent of such a message from its store, and deletes its answers and the answers of its children from the memory.
The user is also able to clear (or even disable) the memory completely in case that behaviour is unwanted.

{
  \looseness = -1
Note that despite what the name ``memory'' might suggest, no fuzzy matching of any sort is done.
At the moment, questions are compared using simple textual identity of their contents.
If the text of a message is slightly different, it will not be considered (this includes renaming of free variables).
This is essentially a trade-off:
the notion of ``fuzziness'' is extremely context-dependent.
For example, for some predicates, a simple change of a constant is meaningless, whereas for other predicates, a conditional rule depends on it.
Designing a reasonable fuzzy matcher is outside of scope here, but an interesting starting point for future work.

}


\section{Message Filtering}
\label{sec:filtering}

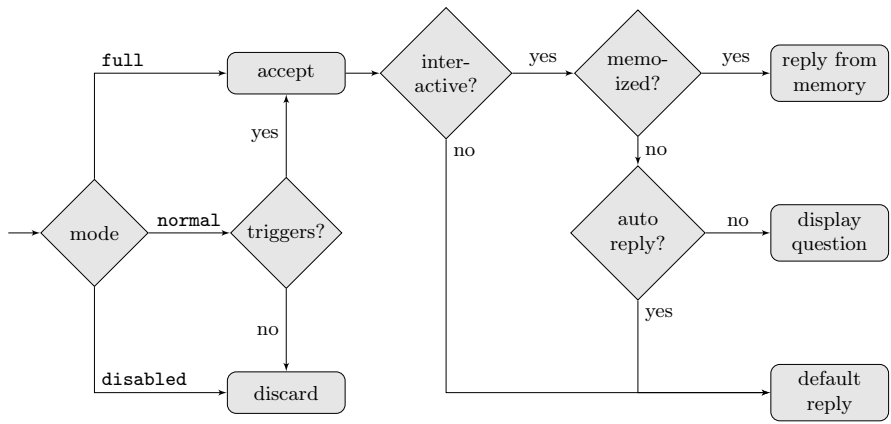
\begin{figure}[t]
  \centering
  \tikzstyle{decision} = [diamond, draw, fill=black!10, text width=4.5em, text badly centered, node distance=3cm, inner sep=0pt]
  \tikzstyle{block} = [rectangle, draw, fill=black!10, text width=5em, text centered, rounded corners, node distance=3cm, minimum height=2em]
  \tikzstyle{line} = [draw, -latex']

  \scalebox{0.85}{
    \begin{tikzpicture}[auto]
      \node [decision] (mode) {mode};
      \path [line] (mode.west) +(-0.5,0) -- (mode.west);

      \node [decision, right of=mode] (triggers) {triggers?};
      \node [block, above of=triggers, node distance=2.5cm] (accept) {accept};
      \node [block, below of=triggers, node distance=2.5cm] (discard) {discard};

      \path [line] (mode) -- node[above] {\texttt{normal}} (triggers);
      \path [line] (triggers) -- node[left] {yes} (accept);
      \path [line] (mode) |- node[above right] {\texttt{full}} (accept);
      \path [line] (triggers) -- node[left] {no} (discard);
      \path [line] (mode) |- node[above right] {\texttt{disabled}} (discard);

      \node [decision, right of=accept, node distance=2.5cm] (interactive) {inter\-active?};
      \node [decision, right of=interactive] (memoized) {memo\-ized?};
      \node [block, right of=memoized] (reply) {reply from memory};

      \path [line] (accept) -- (interactive);
      \path [line] (interactive) -- node[above] {yes} (memoized);
      \path [line] (memoized) -- node[above] {yes} (reply);

      \node [decision, below of=memoized, node distance=2.5cm] (auto) {auto reply?};
      \node [block, right of=auto] (ask) {display question};
      \node [block, below of=ask, node distance=2.5cm] (default) {default reply};

      \path [line] (memoized) -- node[right] {no} (auto);
      \path [line] (auto) -- node[above] {no} (ask);
      \path (auto.south) node[anchor=north west] {yes};
      \path [line] (auto) |- (default);

      \path [line] (interactive) |- (default);
      \path (interactive.south) node[anchor=north west] {no};
    \end{tikzpicture}
  }

  \caption{Message filtering for \texttt{apply} messages}
  \label{fig:filtering}
\end{figure}

Based on the settings described in the previous sections, messages get \emph{filtered}.
The process for \texttt{apply} messages is depicted in Fig. \ref{fig:filtering} and consists of these steps:

\begin{enumerate}
  \item
    Using \emph{normal} verbosity, messages which have not been triggered by a breakpoint are discarded right at the beginning.
    This happens immediately after the simplifier creates them.
    Unless tracing is disabled completely, accepted interactive messages are then transferred to the IDE, where they will be treated as potential questions for the user.
  \item
    If the tracing operates without user intervention (e.g.\ if the user explicitly disabled it earlier), messages are merely logged and answered with a default reply.
    The default reply is chosen so that it does not influence the simplifier in any way, i.e.\ it proceeds as if tracing would be disabled.
  \item
    Some questions are eligible for memoization.
    The memory is queried to check for a match.
  \item
    If \emph{auto reply} is enabled, all remaining questions are also automatically answered with a default reply.
    Otherwise, they are finally being displayed.
    This is scoped to the current focus, i.e.\ only applies to the active questions of the selected command in the proof text.
    A use case for this facility arises when interactive tracing is globally enabled, but the user wants to discharge active questions of some selected commands without having to modify the proof text.
\end{enumerate}

At a first glance, this pipeline might seem a little convoluted.
However, these steps are necessary to match the user's expectation to only get asked if desired, which (ideally) should happen rarely.
Filtering keeps the number of unwanted messages at a minimum.

\section{Related Work}
\label{sec:related}

In this section, we will compare our contributions with the \emph{SWI} implementation of Prolog and the \emph{Maude} rewriting language.
Both systems offer tracing and debugging facilities where the user is able to step through the computation.
The theorem provers Coq, HOL4, HOL Light and PVS have only rudimentary, non-interactive trace facilities, which is why we omit a thorough comparison.

\subsection{Debugging and Tracing in SWI-Prolog}
\emph{Prolog} is a logic programming language \cite{sterling1986art,clocksin2003programming}.
A program consists of a set of \emph{clauses}, namely \emph{rules} and \emph{facts}.
Rules are (usually) Horn clauses, with a head and a body.
Facts are merely rules with an empty body.
Heads may contain variables, and bodies may contain variables not occurring in the head.
Variable names must begin with an upper-case letter or an underscore, whereas \emph{atoms} must begin with a lower-case latter.

\begin{figure}[t]
  \small
  \begin{subfigure}[b]{.5\linewidth}
    \begin{verbatim}
child(a, b).
child(b, c).

descendant(X, X).
descendant(X, Z) :-
  child(X, Y), descendant(Y, Z).
    \end{verbatim}
    \vskip-10pt
    \caption{Input database}
    \label{code:prolog:input}
  \end{subfigure}
  \begin{subfigure}[b]{.6\linewidth}
    \begin{verbatim}
[trace]  ?- descendant(a, c).
   Call: (6) descendant(a, c) ? creep
   Call: (7) child(a, _G1949) ? creep
   Exit: (7) child(a, b) ? creep
   Call: (7) descendant(b, c) ? creep
   Call: (8) child(b, _G1949) ? creep
   Exit: (8) child(b, c) ? creep
   Call: (8) descendant(c, c) ? creep
   Exit: (8) descendant(c, c) ? creep
   Exit: (7) descendant(b, c) ? creep
   Exit: (6) descendant(a, c) ? creep
true ;
   Redo: (8) descendant(c, c) ? creep
   Call: (9) child(c, _G1949) ? creep
   Fail: (9) child(c, _G1949) ? creep
   Fail: (8) descendant(c, c) ? creep
   Fail: (7) descendant(b, c) ? creep
   Fail: (6) descendant(a, c) ? creep
false.
    \end{verbatim}
    \vskip-10pt
    \caption{Query with tracing enabled}
    \label{code:prolog:trace}
  \end{subfigure}

  \caption{Prolog example}
  \label{code:prolog}
\end{figure}

The example in Fig. \ref{code:prolog:input} defines a program with two predicates, \verb~child~ and \verb~descendant~.
A query is basically a predicate with possibly uninstantiated variables, and Prolog tries to instantiate those.
In Prolog terminology, such an expression is a \emph{goal}, and the interpreter attempts to \emph{prove} it.

When proving a goal, Prolog tries to unify the current goal with any of the available clause heads, and proceeds recursively with each item in the body as new subgoals.
This is similar to how the simplifier works in Isabelle:
The left-hand side of a rewrite rule is matched to the current term, and if matches, it tries to solve the preconditions of the rule recursively.

The Prolog implementation \emph{SWI-Prolog} provides a tracing facility for queries \cite[\S\S 2.9,\thinspace 4.38]{fruehwirth2012swi}.
An example for the tracing output can be seen in Fig. \ref{code:prolog:trace} (the term \verb~creep~ denotes continuing the normal process).%
\footnote{A discussion of tracing in Prolog can be found in \cite[\S 8]{clocksin2003programming}, and further analyses in \cite{ducasse1994logic}.
SWI uses a slightly extended variant thereof.}

Apart from continuing the process, SWI offers some additional commands.
The commands relevant for this work are:
\begin{description}
  \item[abort] exits the whole proof attempt
  \item[fail] the current goal is explicitly marked as failure, regardless whether it could have been proved
  \item[ignore] the current goal is explicitly marked as success
  \item[retry] discards the proof of a subgoal and backtracks to the original state of the parent goal
\end{description}

In contrast to SWI, marking a goal as success is not supported in our work.
The simplifier---just like any other tactic---has to justify its steps against Isabelle's proof kernel, which would not accept such a user-declared success.
A possible workaround would be to introduce such theorems ``by cheating'', i.e.\ by explicitly declaring an ``oracle'' which adds an axiom to the proof.
It is an interesting discussion whether or not it is sensible to grant the simplifier the ``privilege'' to generate invalid theorems in tracing mode.

Finally, it is possible to declare breakpoints on predicates.
SWI allows to refine breakpoints with the specific event (referred to as \emph{port}).
For example, the user can specify that they are interested only in \verb~fail~ messages, but not \verb~call~ messages.
However, as soon as such a breakpoint is set, the tracing ceases to be interactive and switches to a log-only mode.%
\footnote{Switching to log-only mode is possible in our work, too.
When in non-interactive mode, trace is only produced for steps matching breakpoints, but no questions are presented to the user.}
In our work, the filtering concept is more sophisticated and allows a fine-grained control over what is being asked.
In SWI, it is not possible to set a breakpoint on terms.

In summary, SWI's features are quite similar to what we have implemented, but differ in a few conceptual points.
First and foremost, the ``execution'' model of Prolog and Isabelle theories differ significantly.
Evaluation of Prolog queries happens sequentially, and any changes in the underlying source code must be explicitly reloaded.
That also means that running queries need to be aborted.
On the other hand, using Isabelle/jEdit, changes in a theory get reloaded immediately and affects pending computations directly.
We can conclude that a memoization mechanism as implemented in our work is not necessary in Prolog.

\subsection{Debugging and Tracing in Maude}

\emph{Maude} is a logic language based on term rewriting \cite{clavel1996principles,clavel2011maude}.
A program consists of data type declarations and equations.
Then, the user can issue a \verb~reduce~ command, which successively applies rewrite rules to an input term.
A short example modelling natural numbers can be seen in Fig. \ref{code:maude:input}.

\begin{figure}[t]
  \small
  \begin{subfigure}[b]{.38\linewidth}
    \begin{verbatim}
fmod SIMPLE-NAT is
  sort Nat .
  op zero : -> Nat .
  op s_ : Nat -> Nat .
  op _+_ : Nat Nat -> Nat .
  op nonzero_ : Nat -> Bool .

  vars N M : Nat .
  ceq nonzero (N + M) = true
    if nonzero N /\ nonzero M .
  eq nonzero (s N) = true .
endfm
    \end{verbatim}
    \caption{A simple program}
    \label{code:maude:input}
  \end{subfigure}
  \hfill
  \begin{subfigure}[b]{.54\linewidth}
    \begin{verbatim}
Maude> reduce nonzero (s zero + s s zero) .
*********** trial #1
ceq nonzero (N + M) = true
  if nonzero N = true /\ nonzero M = true .
N --> s zero
M --> s s zero
*********** solving condition fragment
nonzero N = true
*********** equation
eq nonzero s N = true .
N --> zero
nonzero s zero
--->
true
(...)
*********** success #1
*********** equation
(...)
nonzero (s zero + s s zero)
--->
true
    \end{verbatim}
    \caption{Reducing a term with trace enabled}
    \label{code:maude:trace}
  \end{subfigure}

  \caption{Maude example (based on \cite[\S 2.2]{clavel2011maude})}
  \label{code:maude}
\end{figure}

This snippet declares a data type for natural numbers, along with its two constructors \verb~zero~ and \verb~s~.
Additionally, it defines an addition function, a predicate to check whether a number is non-zero and two equalities for that predicate.
We have left out the equations for addition because they are not needed for the example.

Similar to Isabelle's simplifier, rewrite rules can be \emph{conditional}.
In the trace, it becomes obvious that those are handled exactly like in Isabelle.
A potentially applicable rule gets instantiated with the concrete term, and the preconditions are solved recursively.

The Maude tracing is purely sequential and offers little to no insight into the hierarchical structure when conditional rules are involved.
The trace can be tuned in various ways \cite[\S\S 14.1,\thinspace 18.6]{clavel2011maude}:
for example, Maude allows filtering for named rules and operations (albeit only the outermost operation in the redex is considered).
There is also a wealth of settings which control verbosity of the trace, e.g.\ whether solving preconditions or the definition of rules should be included in the trace.

Apart from controlling the textual output, it is also possible to enable output colouring, similarly to the highlighting in Isabelle.
The major difference here is that Maude distinguishes between \emph{constructor} and \emph{nonconstructor} symbols, whereas such a distinction is not made in Isabelle.
An indicator for problems in the set of rewrite rules of a Maude program is when a term does not get fully rewritten, which is defined as nonconstructor symbols still occur after reducing \cite[\S 14.1.2]{clavel2011maude}.
Hence, colouring symbols differently greatly improves debugging experience in Maude, because it also gives hints into when exactly a problem has been introduced.

In addition to tracing, there is also a debugging facility.
It can be configured with breakpoints in the same way as the tracing.
When a breakpoint is hit, the user can resume or abort the whole computation, but also (on request) observe the stack of recursive invocations.
The latter also includes a textual explanation, e.g.\ that the current term is being rewritten to solve a precondition.

A distinguishing feature of the debugger is that it allows to execute a new \verb~reduce~ command when a debugging session is active.
This allows the user to quickly check related terms and hence better understand an issue with the original term.

Maude has been an active target for research for refining the trace even further, providing insights into when and how a particular term emerged during reducing (e.g.\ Alpuente et.al.\ \cite{alpuente2013slicing}).
Term provenance could certainly be an interesting extension for our work on the simplifier trace, but would require significantly more instrumentation of the simplifier.

\section{Evaluation and Future Work}
\label{sec:eval}

In this section, we will briefly discuss the performance of the new tracing and the practical usability.
Furthermore, possibilities for future work are explored.

\subsection{Performance}

\begin{table}[t]
  \centering
  \caption{Execution times to simplify $10^x \cdot 10^y$ (in seconds, without rendering)}
  \begin{tabular}{|c|c|r|r|r|r|}
    \hline
    $\mathbf x$ & $\mathbf y$ & disabled & old tracing & \texttt{normal} & \texttt{full}
    \\\hline\hline
    $10$ & $10$ & $0.02$ & $0.56$ & $0.39$ & $0.92$ \\\hline
    $20$ & $10$ & $0.04$ & $1.61$ & $1.21$ & $2.98$ \\\hline
    $20$ & $20$ & $0.08$ & $3.15$ & $2.44$ & $6.00$ \\\hline
  \end{tabular}
  \label{tab:eval-perf}
\end{table}

Logging the simplification process obviously incurs a measurable overhead.
For example, consider the expression scheme $10^x \cdot 10^y$ (which will be evaluated to a literal number by the simplifier given concrete values of $x$ and $y$).
The test machine is an Intel Core \mbox{i7-3540M} with a peak frequency of about \SI{3.7}{\GHz}, and \SI{8}{\gibi\byte} of memory.
Execution times have been collected using the \verb~Timing~ panel in Isabelle/jEdit without having the IDE render the trace data.
The results can be seen in Table \ref{tab:eval-perf}.

As can be seen in the table, the simplifier itself is pretty fast, but enabling the trace slows down the process significantly.
Note that for measuring in \verb~normal~ mode, no breakpoints have been set, hence these numbers show just the overhead of the tracing:
In every single rewrite step, the tracing hooks have to be called and various lookups are performed.
The most interesting comparison is between the old tracing and \verb~full~ mode, where the ratio is roughly $1:2$.
This can be explained by the fact that the full mode collects more information and processes it more thoroughly than the old tracing.

The slowest component overall is the GUI though (times not shown in the table), which requires about \SI{5}{\second} to display the full trace for $x = y = 10$.
The old tracing needs just \SI{1}{\second}.
The difference here can again be explained by the amount of information processed; in particular, the old tracing does not have to reconstruct the hierarchical structure of the trace messages.
For $x = 20, y = 10$, the GUI already needs \SI{37}{\second} to render the full trace, resulting in roughly \num{200000} lines.
However---once rendered---scrolling, collapsing and expanding nodes is instantaneous.
Hence, it is generally advisable to enable the simplifier trace only if necessary when dealing with huge traces.
For smaller traces which are in the order of hundreds of messages, e.g.\ the example from the introduction, all GUI actions are instantaneous.

\subsection{Future Work}

There are multiple dimensions in which this work can be extended in the future.

Integration into the Isabelle system would benefit by adapting more tactics to use the new tracing mechanisms, since many of them can be modelled in a hierarchical manner, and more message types could be introduced.
For example, the simplifier supports custom simplification procedures (``ML functions that prove rewrite rules on demand'', \cite[\S 9.3.5]{wenzel2013reference}).
In the new tracing, invocations of those ``simprocs'' are ignored.
A future extension could be to provide hook functions to those simprocs.
Besides rewriting, proof reconstruction tactics \cite{blanchette2011automatic} fit naturally into the recursive model; for example the \texttt{metis} resolution prover to replay proofs from external ATPs (automated theorem provers).

The user experience could be improved by asking even less questions (introduce fuzzy matching in the memoization) or providing more information per question (e.g.\ the whole term context of the redex).
Interesting context data includes for example term provenance, i.e.\ tracking how a particular subterm in the result emerged during the rewriting process.
Also, there still some oddities arising from the asynchronous document model which leads to undesired delays in the user interface under certain circumstances.
Resolving this would most likely require nontrivial changes in the message passing mechanisms in Isabelle/Scala.

Furthermore, the current implementation would hang indefinitely if the simplifier fails to terminate.
The old tracing in recent versions of Isabelle pauses the simplification process after a certain number of messages have been emitted and waits for user confirmation to continue.
(Previously, the IDE could easily be rendered unresponsive when the memory pressure became too high.)
In the new tracing, this problem occurs less often, because---if running in \texttt{normal} mode---most messages are discarded, so users will most likely see no messages.
This is not a big problem, since non-termination may happen in other commands in Isabelle as well, and the user can transparently cancel such invocations just by deleting the corresponding text in the theory source.
To allow for better debugging, we could additionally perform some basic termination checks, along with a new message type informing the user of the problem.

Support for other Isabelle IDEs is currently not supported.
IDEs based on the document model (e.g.\ Isabelle/Eclipse\footnote{developed by Andrius Velykis, \url{http://andrius.velykis.lt/research/}}) would require reimplementing the pure GUI parts of the tracing.
As for Proof General \cite{aspinall2000proof}, all Scala parts of this work would need to be completely reimplemented, possibly in a different programming language.

\subsection{Case Study: A Parallelized Simplifier}

As indicated earlier, because of the loose coupling between simplifier and tracing, a parallelized simplifier can easily be supported.
Since Isabelle's simplifier is quite complex, we implemented an extremely stripped-down but parallel version in order to test this capability.
The ``proof of concept'' simplifier is not nearly as powerful as the ``real'' one---given that it consists only of roughly 100 lines of code---but it splits off some of the work into parallel tasks:
Preconditions of a conditinal rewrite rule are resolved simultaneously, and if multiple rewrite rules are applicable, they are tried in parallel.

The implementation overhead of parallelization is low, because Isabelle/ML offers library support for both concurrent and parallel programming.
The primary abstraction are \emph{future} values \cite{matthews2010efficient,wenzel2009parallel}:
the type constructor \verb~'a future~ denotes a value of type \verb~'a~ becoming available at a later time.
Parallelizing existing purely (i.e.\ not side-effecting) functional code is simple, because it only requires to wrap subexpressions into the \verb~fork~ constructor of futures, which automatically evaluates them on a pool of worker threads.
There are combinators for combining parallel computations and to wait on the completion of the result.

A little bit more intervention is required when parallelizing side-effecting code.
However, since the Isabelle document model is asynchronous by its nature, it already offers message-oriented ways for communication between the prover and the IDE process.
Consequently, the system already deals with common concurrency issues; in this case the only remaining task was to ensure that no deadlocks occur.

For the simple examples we tried, a parallelized simplifier did not yield a significant speed-up compared to the sequential simplifier.
However, the key insight lies in the user interaction.
When multiple preconditions are solved simultaneously, the user might see more than one active question at the same time which could potentially be confusing.
Furthermore, race conditions could lead to nondeterministic traces, a problem well-known when debugging parallel imperative programs.
Hence, the question of a reasonable user interface for parallel simplification remains open.

\section{Conclusion}

We presented a generic tracing facility for Isabelle tactics which replaces the old simplifier trace.
That new facility is interactive, highly configurable, and intuitive to operate.
The performance and implementation impact on the rest of the system turned out to be rather small.
Nonetheless, it became possible to provide more insights for the user into the simplification process.

The design goal that the amount of interaction with the user should be kept low has been achieved.
Various sophisticated filtering and memoization techniques help maintaining a good trade-off between flexibility and opacity of the system.

The new tracing mechanism is available in the main Isabelle development repository since early February 2014.

\paragraph{Acknowledgements} 
I thank Tobias Nipkow and Lars Noschinski for encouraging me to implement a new simplifier trace.
I am grateful to Makarius Wenzel who commented multiple times on various stages of the code and patiently answered my questions about internals of the Isabelle system.
Dmitriy Traytel and Cornelius Diekmann commented on early drafts of this paper.

\bibliography{references}

\begin{thebibliography}{10}
\providecommand{\url}[1]{\texttt{#1}}
\providecommand{\urlprefix}{URL }

\bibitem{alpuente2013slicing}
Alpuente, M., Ballis, D., Frechina, F., Sapi{\~n}a, J.: Slicing-{B}ased {T}race
  {A}nalysis of {R}ewriting {L}ogic {S}pecifications with i{J}ulienne. In:
  ESOP. pp. 121--124 (2013)

\bibitem{aspinall2000proof}
Aspinall, D.: Proof {G}eneral: {A} generic tool for proof development. In:
  Tools and Algorithms for the Construction and Analysis of Systems, pp.
  38--43. Springer (2000)

\bibitem{blanchette2011automatic}
Blanchette, J.C., Bulwahn, L., Nipkow, T.: {Automatic Proof and Disproof in
  Isabelle/HOL}. In: Frontiers of Combining Systems. pp. 12--27. Springer
  (2011)

\bibitem{clavel2011maude}
Clavel, M., Dur{\'a}n, F., Eker, S., Lincoln, P., Mart{\'i}-Oliet, N.,
  Meseguer, J., Talcott, C.: Maude manual (version 2.6)

\bibitem{clavel1996principles}
Clavel, M., Eker, S., Lincoln, P., Meseguer, J.: Principles of {Maude}. In:
  Meseguer, J. (ed.) Electronic Notes in Theoretical Computer Science. vol.~4.
  Elsevier Science Publishers (1996)

\bibitem{clocksin2003programming}
Clocksin, W.F., Mellish, C.S.: Programming in {P}rolog: {U}sing the {ISO}
  standard. Springer (2003)

\bibitem{ducasse1994logic}
Ducass{\'e}, M., Noy{\'e}, J.: Logic programming environments: {D}ynamic
  program analysis and debugging. The Journal of Logic Programming  19--20,
  Supplement 1,  351--384 (1994)

\bibitem{fruehwirth2012swi}
Fruehwirth, T., Wielemaker, J., De~Koninck, L.: {SWI} {P}rolog Reference Manual
  6.2.2. Books on Demand (2012)

\bibitem{matthews2010efficient}
Matthews, D.C., Wenzel, M.: Efficient parallel programming in {P}oly/{ML} and
  {I}sabelle/{ML}. In: Proceedings of the 5th ACM SIGPLAN workshop on
  Declarative aspects of multicore programming. pp. 53--62. ACM (2010)

\bibitem{paulson2002isabelle}
Nipkow, T., Paulson, L.C., Wenzel, M.: Isabelle/{HOL}: a proof assistant for
  higher-order logic, Lecture Notes in Computer Science, vol. 2283. Springer
  (2002)

\bibitem{sterling1986art}
Sterling, L., Shapiro, E.Y.: The {A}rt of {P}rolog: {A}dvanced {P}rogramming
  {T}echniques. MIT Press Cambridge (1994)

\bibitem{gdb}
{The GNU Project}: {GDB: The GNU Project Debugger},
  \url{https://www.gnu.org/software/gdb/}

\bibitem{wenzel2009parallel}
Wenzel, M.: Parallel proof checking in {I}sabelle/{I}sar. In: Proceedings of
  the ACM SIGSAM 2009 International Workshop on Programming Languages for
  Mechanized Mathematics Systems. pp. 13--29. ACM (2009)

\bibitem{wenzel2012asynchronous}
Wenzel, M.: Asynchronous proof processing with {I}sabelle/{S}cala and
  {I}sabelle/j{E}dit. Electronic Notes in Theoretical Computer Science  285,
  101--114 (2012)

\bibitem{wenzel2012isabelle}
Wenzel, M.: Isabelle/j{E}dit -- a {P}rover {IDE} within the {PIDE} framework.
  In: Intelligent Computer Mathematics, pp. 468--471. Springer (2012)

\bibitem{wenzel2013reference}
Wenzel, M.: {The Isabelle/Isar Reference Manual} (2013)

\end{thebibliography}
\bibliographystyle{splncs03}

\end{document}